\documentstyle[version2,preprint,aps]{revtex}
\draft
\begin{document}
\columnsep -.375in
\begin{title}
Kosterlitz-Thouless Transition and Short Range Spatial Correlations\\
in an Extended Hubbard Model
\end{title}
 
\author{Qimiao Si and J. Lleweilun Smith}
 
\begin{instit}
Department of Physics, Rice University, Houston, TX 77251-1892, USA
\end{instit}
 
\begin{instit}
\end{instit} 
 
\begin{abstract}
We study the competition between intersite and local correlations
in a spinless two-band extended Hubbard model by taking an
alternative limit of infinite dimensions. We find that
the intersite density fluctuations suppress the charge Kondo 
energy scale and lead to a Fermi liquid to non-Fermi liquid 
transition for repulsive on-site density-density interactions.
In the absence of intersite interactions, this transition 
reduces to the known Kosterlitz-Thouless transition.
We show that a new line of non-Fermi liquid fixed points
replace those of the zero intersite interaction problem.
\end{abstract}
 
\vskip 0.2 in
 

The two-band extended Hubbard model is a realistic starting point
both for the high $T_c$ cuprates and for many heavy fermion 
systems. The model contains a strongly correlated band
and a weakly correlated one. At the phenomenological level,
the low energy properties of the conventional heavy fermions
(such as $CeCu_6$ and $UPt_3$) are well described by the Fermi
liquid theory\cite{Steglich}, while those of certain novel
$f-$electron materials\cite{Maple} and the high $T_c$ cuprates
appear not. The theoretical question, then, is: under what conditions 
do electron correlations lead to a non-Fermi liquid in this model?

Recently, some progress has been made on the understanding of
this model\cite{SK,Perakis}. In the large dimensionality ($D$)
limit, the local density-density interactions alone are found
to cause Kosterlitz-Thouless\cite{KT} type quantum phase
transitions from a Fermi liquid to non-Fermi liquid metallic 
states. In the spinful case, the resulting non-Fermi liquids
have spin-charge separation\cite{SK,KS}. The large $D$
limit\cite{Georges} has the advantage that local correlations
are treated in a dynamical fashion, but the disadvantage that
all intersite interactions reduce to Hartree-Fock terms; 
{\it no spatial fluctuations survive}. For physical systems in
finite dimensions, intersite RKKY or Superexchange type
interactions are expected to compete with local
correlations\cite{Doniach,Jones}.
Unfortunately, for ``paramagnetic'' phases far away
from spatial ordering transitions, there exists no
controlled theoretical method to address such a competition.
From the large $D$ point of view, one way to recover spatial
fluctuations is the perturbative ${1 / D}$ expansion\cite{Schiller}.
The practicality of this procedure is unclear at this stage.
An alternative route is a loop expansion with the requirement that
the $D=\infty$ results be recovered at the saddle-point level,
as has been constructed in models with certain forms of quenched
disorder\cite{Vlad}. For clean systems, this turns out to be
difficult to formulate.

In this paper, we take an alternative large $D$ limit to
study the interplay between local correlations and short range
spatial fluctuations in the two band extended Hubbard model.
We introduce an explicit intersite density-density interaction
term and scale the interaction strength in terms of the
dimensionality such that its {\it fluctuation part} survives
the large $D$ limit. This procedure leads to an impurity
embedded in a self-consistent fermionic bath {\it and a 
self-consistent bosonic bath}. We will study primarily the
spinless version of the model for which asymptotically exact
results can be derived analytically. The results are of
direct relevance to the charge sector of the spin-charge-separated
intermediate phase of the spinful extended Hubbard model\cite{SK,KS}.

The model is defined by the following Hamiltonian,

\begin{eqnarray}
H = &&\sum_{i} [\epsilon_d^0 n_{d_i}  + t (c_i^{\dagger}d_i + h.c.) 
+ V :n_{c_i}: ~ :n_{d_i}:]\nonumber\\
&&+\sum_{<ij>} [ t_{ij}c_i^{\dagger}c_j +
(v_{ij}/2):n_{d_i}:~:n_{d_j}: ]
\label{hamiltonian}
\end{eqnarray}
The spinless $d-$ electrons are dispersionless, with an
energy level $\epsilon^o_d$. 
The spinless $c-$ electrons have a hopping matrix $t_{ij}$.
$<ij>$ labels the nearest-neighbor sites. 
The $t-$term describes hybridization, and the $V-$term the
on-site density-density interaction. The $v_{ij}-$ term
is an intersite ``charge RKKY'' repulsive interaction.
The standard large $D$ limit\cite{Georges} is taken with 
$t_{<ij>}$ scaled to be of order ${1 / \sqrt{2D}}$ and 
$v_{<ij>}$ of order ${1 / D}$. In this limit, only the static
part of the $v_{ij}-$term survives. Here, we scale
$v_{<ij>}=\frac{v_0}{\sqrt{2D}}$ and take the large
$D$ limit keeping $v_0$ fixed. As shown below, this limit is well
defined when we retain only the dynamical density modes.
We achieve this through normal ordering, $:n: \equiv n - <n>$.
We focus on states without long range ordering, for which the 
Hartree terms can be absorbed by the chemical potential.

We first give a general formulation of this alternative large
$D$ limit. Following the standard procedure, we divide the 
Hamiltonian into local and intersite parts,

\begin{eqnarray}
H = \sum_i h_i + \sum_{<ij>} ( h_{t,ij} + h_{v,ij})
\end{eqnarray}
where $h_i$ is on-site, $h_{t,ij} = t_{ij} \psi_i^{\dagger}\psi_j$,
and $ h_{v,ij} = (v_{ij}/2) :n_{\psi_i}: :n_{\psi_j}: $.
For generality, we develop the formalism with
the $\psi_i^{\dagger}$ fields for possible multi- bands or 
spin-components; for the Hamiltonian 
(\ref{hamiltonian}), $h_{t,ij} = t_{ij}
c_i^{\dagger}c_j$, and $ h_{v,ij} = (v_{ij}/2) :n_{d_i}: :n_{d_j}: $.
Within a path-integral representation, we divide the action for the
lattice model into $S = S_0 + S^{(0)} + \Delta S $, where $S_0$,
$S^{(0)}$, and $\Delta S$ are the actions associated with
$h_0$, $\sum_{i \ne 0} h_i + \sum_{<ij \ne 0>} 
(h_{t,ij} +h_{v,ij})$, and $\sum_{i}( h_{t,0i} + h_{v,0i} + H.c.)$,
respectively. Integrating out all the degrees of freedom except
at site 0 leads to the following effective action,
$S^{eff} = S_0 - \sum^{\infty}_{n=1} (-1)^n 
\frac{<\Delta S^n>^{(0)}_c}{n!}$,
where $<>_c^{(0)}$ denotes connected correlation functions
in terms of $S^{(0)}$.
$<\Delta S^n>^{(0)}_c$ modifies the on-site action with terms that
involve $n$ site-0 density/fermion modes. The coefficients
of these operators are retarded and are given, for each $n$,
by $n$-operator correlation functions with respect to $S^{(0)}$.
The order in ${1 /D}$ of each term can be 
determined through a cumulant expansion of the original 
lattice Hamiltonian\cite{Metzner}.
We find that, to the leading order in ${1 / D}$, no interference
between $h_{t,ij}$ and $h_{v,ij}$ terms is allowed, except in 
local decorations. 
This absence of interference implies that,
n-point correlation functions have the usual dependence on ${1 /D}$.
As a result, for all $n>2$, $<\Delta S^n>^{(0)}_c$ vanishes
as $D \rightarrow \infty$\cite{Georges}.
The absence of interference also leads to separate Dyson equations
for $\chi_{ij}$ and $G_{ij}$ in terms of their respective effective
cumulants. This in turn implies that,
$G_{ij} = G_{ii} G'_{ij} G_{jj}$,
$\chi_{ij} = \chi_{ii} \chi'_{ij} \chi_{jj}$,
$G_{ij}^{(0)} = G_{ij} - G_{ii}G'_{i0} 
G_{00}G'_{0j}G_{jj}$, and
$\chi_{ij}^{(0)} = \chi_{ij} - \chi_{ii} \chi'_{i0} 
\chi_{00}\chi'_{0j}\chi_{jj}$,
where $G'_{ij} 
\equiv \sum_{paths}t_{il_1}G_{l_1l_1}t_{l_1l_2}G_{l_2l_2} ...
G_{l_n l_n}t_{l_n j} $,
$\chi'_{ij} \equiv \sum_{paths} v_{il_1}\chi_{l_1l_1}
v_{l_1l_2}\chi_{l_2l_2} ...\chi_{l_n l_n} v_{l_n j} $,
and $[i,l_1,l_2, ...,l_n,j]$ labels a non-self-retracing path
from site $i$ to site $j$. 
It then follows that,
\begin{eqnarray}
G_{ij}^{(0)} &&= G_{ij} - G_{i0}G_{0j}/{G_{00}} \nonumber\\
\chi_{ij}^{(0)} &&= \chi_{ij} - \chi_{i0} \chi_{0j}/{\chi_{00}}
\label{gchi.rest}
\end{eqnarray}
From Eq. (\ref{gchi.rest}), $<:n_i:>^{(0)} \equiv  <n_i>^{(0)} 
- <n_i>$ is of order ${1 / D}$. Hence,  in the large
$D$ limit, $<\Delta S>^{(0)}$ also vanishes; only
$<\Delta S^2>^{(0)}_c$ survives, leading to

\begin{eqnarray}
S^{eff} =  -\int_0^{\beta}&& d\tau \int_0^{\beta} d\tau' 
[\psi^{\dagger}(\tau) g_0^{-1} (\tau-\tau') \psi (\tau')
\nonumber\\
&&+:n_{\psi}:(\tau) \chi_0^{-1} (\tau - \tau ' ):n_{\psi}:(\tau ')]
+ S_0 
\label{seff}
\end{eqnarray}
where 

\begin{eqnarray}
g_{0}^{-1}(i\omega_n) &&= -
\sum_{ij}t_{i0}t_{0j}G_{ij}^{(0)}(i \omega_n) \nonumber\\
\chi_{0}^{-1}(i \nu_n) &&= \sum_{ij}v_{i0}v_{0j} \chi_{ij}^{(0)}(i \nu_n)
\label{selfconsistent}
\end{eqnarray}
The Dyson equations of the lattice Hamiltonian implies that
$G_{ij}$ is still determined by the on-site self-energy
alone; $\chi_{ij}$ is determined by the on-site effective
cumulant, or, equivalently, the on-site part of the vertex function 
irreducible in terms of both the particle hole bubble and
the single $v_{ij}$ line. All these local quantities
can be calculated directly in terms of $S^{eff}$.
Therefore, given an $S^{eff}$, we can calculate all
$G_{ij}$ and $\chi_{ij}$ and hence, via.
Eq. (\ref{selfconsistent}), $g_{0}^{-1}(i\omega_n)$ and 
$\chi_{0}^{-1}(i \nu_n)$. The self-consistent equations
(\ref{seff},\ref{selfconsistent},\ref{gchi.rest})
indeed close. They define the dynamical mean field
equations associated with our alternative large $D$ limit.
We note that related mean field equations arise in the metallic
spin-glass problem\cite{Anirvan}.
As usual, the retarded $g_0^{-1}$ can be represented 
in terms of a self-consistent non-interacting fermionic bath\cite{Georges}.
Similarly, an additional non-interacting {\it bosonic} bath can
be used to represent the retarded $\chi_0^{-1}$ term. We have
therefore a self-consistent Anderson-like impurity model 
coupled to screening bosons\cite{Haldane}.

We now apply this formalism to
the extended Hubbard model (\ref{hamiltonian}).
The self-consistent impurity model is a resonant-level
model with an additional bosonic bath,

\begin{eqnarray}
H^{eff}_{imp}=&&E_d^0d_{0}^{\dagger}d_{0} + H_0
+t(c_{0}^{\dagger}d_{0}+h.c.) + V:n_{c_0}: :n_{d_0}: \nonumber\\
&& + \sum_{q}W_{q}\rho_{q}^{\dagger}\rho_{q} 
+ \sum_{q}F_{q}:n_{d_0}: (\rho_{q} + \rho_{q}^{\dagger})
\label{heff}
\end{eqnarray}
where $E_d^0 = \epsilon_d^0 - \mu$, and 
$H_0$ describes the local $c_0$ electron coupled to a 
non-interacting fermionic bath whose dispersion,
together with the parameters associated with the bosonic bath,
$W_q$ and $F_q$, are determined from the self-consistency
eqs. (\ref{selfconsistent}).
In the remainder of this paper, we focus on taking the
large $D$ limit with a Bethe lattice. This has the
advantage that the bare density of states is bounded.
For $v_0=0$, it was shown both analytically\cite{SK} and 
numerically\cite{SRKR} that
the density of states of the self-consistent fermionic bath
is non-singular in the metallic regime.
We find this continues to be the case for finite $v_0$.
This then allows an asymptotically exact analytic analysis
of our self-consistent problem.
Note that the spectral function of the
bosonic bath is arbitrary (and is in 
fact non-ohmic in the non-Fermi liquid case, see below).

The asymptotically exact analysis is carried out by writing
the partition function of $H^{eff}_{imp}$ in a kink-gas
representation. The procedure parallels what was 
detailed in Ref. \cite{SK}. 
We only note that, the effect of the additional
bosonic bath on the action of a particular history can be
treated by a time-dependent canonical transformation, for 
{\it arbitrary form of the boson spectral function}.
The resulting kink-gas action is,

\begin{eqnarray}
S(\tau_{2n}, &&...,\tau_{1})= - 2n ~ln(y) 
+\sum_i (-1)^i ~h ~({\tau_{i+1} - \tau_i})/\xi_0
\nonumber\\
&& + \sum_{i < j}(-1)^{i+j} [ 2\epsilon 
~ln ({\tau_{j}-\tau_{i}})/{\xi_{0}} 
+ K({\tau_{j}-\tau_{i}}) ]
\label{kinkgas}
\end{eqnarray}
Here, $[\tau_{2n},...,\tau_{1}]$, for $n=1, 2, ...$, labels a sequence
of kink events along the imaginary time axis. The fugacity
is given by $y=t\rho_c$, where $\rho_c$ is the density of states
of the self-consistent fermionic bath at the Fermi energy. 
$\epsilon= (1/2) [1 - (1/\pi)(
{tan^{-1}(\pi \rho_{c} (1-n_d)V)}
+{tan^{-1}(\pi \rho_{c} n_dV)})]^{2}$ is the stiffness constant. 
The logarithmic interaction among the kinks is mediated by
the fermionic bath. $h = E_d^0 \xi_0$ is the symmetry breaking 
field. Finally, $K(\tau)$ describes an additional long-range 
interaction of the kinks induced by the bosonic bath.
Due to the self-consistency, $K(\tau )$ is determined entirely
by the full local susceptibility,

\begin{eqnarray}
{\partial^{2}K(\tau)}/{\partial \tau^{2}}= v_0^2 \chi(\tau) 
\label{K.selfconsistent}
\end{eqnarray}

Anticipating the possible ${1 / \tau^{\alpha}}$ form for the
local susceptibility, we first solve the kink-gas action 
(\ref{kinkgas}) with 

\begin{eqnarray}
K(\tau) = \lambda [({\tau / \xi_0})^{2-\alpha} - 1]/(2-\alpha) 
\label{K.fixedalpha}
\end{eqnarray}
for a fixed $\alpha$ value. We will consider the case
with vanishing renormalized symmetry breaking field, $h^* $.
We have derived the RG equations for $\alpha$ close to $2$,

\begin{eqnarray}
d y / {d ln \xi}~&&=~ y[ 1 - (\epsilon + {\lambda / 2})]\nonumber\\
{d \epsilon / {d ln \xi}}~&&=~ -4 \epsilon y^2 \nonumber\\
{d \lambda / {d ln \xi}}~&&=~ \lambda [(2-\alpha) - 4 y^2 ] \nonumber\\
{d h / {d ln \xi}}~&&=~ h ( 1 - 2 y^2)
\label{scaling}
\end{eqnarray}
valid for small $h$. The RG flow diagram is given in Figure 1.
The flow in the $y-\epsilon$ plane (the dotted lines) describes a
Kosterlitz-Thouless transition. It goes
to a strong coupling fixed point when 
$\epsilon < \epsilon^{crit}=1$, and to a line
of weak coupling fixed points when $\epsilon > \epsilon^{crit}$.
The flow in the $y-\lambda$ plane (the dashed lines) are
those of the Ising model with a long range 
${1 / \tau^{\alpha}}$ interaction\cite{Kosterlitz,Dyson}.
There exists an unstable fixed point at $(y^*, \lambda^*)=
(\sqrt{2-\alpha}/2, 2)$, which describes a second 
order phase transition. Close to the origin, the separatrix has 
the form $\lambda^{sep} \approx 2 [y/(2-\alpha )]^{2-\alpha}$. 
The flow goes to a strong coupling fixed point for 
$\lambda < \lambda^{sep}$, and to a weak coupling fixed 
point when $\lambda < \lambda^{sep}$. In the language of the
long-range Ising model, $\alpha = 2$ is the lower critical range,
and $\alpha = 3/2$ the upper critical range\cite{Kosterlitz}.
The RG flows in between these two planes, denoted by the solid lines,
interpolate between these two limits. There exist a line of weak coupling
fixed points with $y^*=0$, $\lambda^*=\infty$, and a finite $\epsilon^*$.

These results can be used to solve our self-consistent problem.
We focus on the mixed valence regime only. We specify the phase diagram
in terms of the three-dimensional parameter space spanned by
$g_t = t\rho_0$, $g_V = [1 - (2/\pi){tan^{-1}(\pi \rho_{0} V/2)}]$,
and $g_v = \rho_0 v_0$, where $\rho_0$ is the bare density of
states of the conduction electrons at the Fermi energy.
The Kosterlitz-Thouless 
transition at $g_v=0$ describes the charge-Kondo effect
\cite{Wiegmann}. The critical value $g_V^{crit}$ corresponds to
an attractive $V^{crit}$ such that $\rho_{c} V^{crit} =-(2 / \pi) 
tan[(\sqrt{2} - 1)\pi/2] $.
When $g_V < g_V^{crit}$, i.e., $-V < V^{crit}_0$, the solution
is a Fermi liquid with the usual form for the local density
susceptibility,

\begin{eqnarray}
\chi (\tau ) \sim ( {1 / \Delta^* \tau } )^2~~~~~~~~~~~~~~ 
{\rm for} ~~~~~~\tau >> {1 / \Delta^*}
\label{susc.fl}
\end{eqnarray}
where $\Delta^*$ denotes the charge-Kondo energy scale, which
acts as the renormalized Fermi energy.
As $V$ approaches $V_c$ from the Fermi liquid side, $\Delta^*$
vanishes in the Kosterlitz-Thouless fashion,
$\Delta^ * \approx ( \rho_c )^{-1} \exp [-1/ \sqrt{\epsilon^{crit} 
- \epsilon}]$.
For $g_V > g_V^{crit}$, i.e., $-V > V^{crit}_0$, the solution is a 
line of non-Fermi liquids with the connected local density susceptibility,
\begin{eqnarray}
\chi (\tau) \approx ( {1 / \rho_c \tau } )^{\alpha}
\label{susc.nfl}
\end{eqnarray}
The exponent $\alpha$ is interaction-dependent, increasing
from 0 to 2 as one moves away from the critical
point\cite{Bhatt,Imbrie}.

The intersite interaction $v_0$ modifies the phase diagram in several
ways. Consider first $g_V > g_V^{crit}$. The line of fixed points of
the $v_0=0$ problem becomes unstable. In terms of the
parameters appropriate for the kink-gas action
Eq. (\ref{kinkgas}), the RG flow is towards an infinite value of
the $\lambda$ coupling. Usually, one cannot specify the
resulting fixed points when a coupling constant flows towards infinity.
Remarkably, in our case we can determine the correlation functions in 
the new fixed points due to the special feature of the solution 
to the kink-gas action 
Eqs. (\ref{kinkgas}, \ref{K.fixedalpha}):
the local susceptibility has an algebraic time dependence 
with an exponent identical to that of the range of
$K(\tau)$\cite{strong,Imbrie}. The RG flow is towards another
line of fixed points with an infinite $\lambda^*$, a finite 
$\epsilon^*$, and a vanishing hybridization $y^*$. The connected
local susceptibility remains to have the algebraic
${1 / \tau^{\alpha^*}}$ form, with the exponent $\alpha^*$
entirely determined by $\epsilon^*$.

For $g_V<g_V^{crit}$, we are able to establish
the existence of a phase transition as $v_0$ is increased. 
First, the solution must be a non-Fermi liquid
for sufficiently strong $v_o$. Suppose that the 
solution is a Fermi liquid, with a renormalized Fermi energy
$E^*$. Then the local susceptibility has a long time
$(1 / E^*\tau )^2$ dependence for $\tau$ longer than
$1 / E^*$. At times up to $1 / E^*$, if the
local susceptibility decays slower than ${1 / \tau^2}$,
then the corresponding $K(\tau)$ in this intermediate
time range has the form Eq. (\ref{K.fixedalpha}) with $\alpha < 2$ 
and $\lambda = v_0^2A$, where $A$ is the prefactor in the decay of
the local susceptibility. The scaling equations (\ref{scaling})
imply that, as long as $\epsilon + \lambda /2 >1 $, the fugacity 
decreases in this time range. If the local susceptibility
decays faster than ${1 / \tau^2}$, one can still choose a $v_0$ 
sufficiently large such that the fugacity does not increase up to the
time scale $1 / E^*$. In both cases, at times beyond
$1 / E^*$, the kink-gas action corresponds to a Coulomb gas
with a stiffness constant $\epsilon'=\epsilon (\xi={1 / E^*})+  
({v_0 / {\sqrt{2} E^*}} )^2$ and a small 
value of fugacity $y'$. When $v_0$ becomes sufficiently large, 
$( \epsilon ' , y')$ will lie in the weak coupling side of the
RG flow, making $E^* =0$. Therefore, the Fermi liquid 
solution cannot occur.  The only self-consistent solution in this
large $v_0$ regime is one similar to what we found for the
$g_V > g_V^{crit}$ case, characterized by a $y^*=0$, a finite
$\epsilon^*$, and an infinite $\lambda^*$. The local susceptibility
has the form of Eq. (\ref{susc.nfl}), with the exponent being
smaller than two. Physically, the intersite density-density
interactions provide charge-screening,
which contribute to the orthogonality
effect\cite{Perakis,Haldane}. In the mixed valence regime,
this orthogonality helps realize the weak coupling fixed point
with incoherent charge excitations.

For sufficiently small $v_0$, on the other hand,
the Fermi liquid solution is stable. This can be seen by a 
$v_0-$expansion around the $v_0=0$ solution. We replace
$K(\tau)$ in the kink-gas action Eq. (\ref{kinkgas}) by
what we would get from Eq. (\ref{K.selfconsistent}) with
the local susceptibility $\chi(\tau)$ of the $v_0=0$ problem,
Eq. (\ref{susc.fl}). At the time
scale $\xi ' = {1 / \Delta^*}$, $\epsilon(\xi ') < \epsilon (\xi_0)
< 1$, and $K(\tau) = (v_0/\Delta^*)^2  \ln (\tau / \xi ')$. 
From Eqs. (\ref{K.selfconsistent}), the kink-gas action
at $\xi > \xi'$ is a Coulomb gas with a stiffness
constant $\epsilon ' = \epsilon(\xi ') + (v_0 /\sqrt{2}\Delta^*)^2$.
For sufficiently small $v_0$, $\epsilon ' < \epsilon^{crit}=1$.
In this range, the Fermi liquid solution is self-consistent.
Self-consistency, however, will modify the renormalized 
Fermi energy, making it unlikely that the phase transition
is of the Kosterlitz-Thouless type. The precise nature of the
transition is beyond the reach of our RG formalism.
The schematic phase diagram is given in Fig. 2.

As a result, non-Fermi liquids with self-similar 
correlation functions occur even for repulsive values of
the on-site density-density interaction. 

In the non-Fermi liquid case, that the renormalized $\lambda^*$ 
is infinite is one indication that the ground state cannot be the 
``paramagnetic'' metallic state. The fact that the Ising model 
corresponding to the kink-gas action Eqs. (\ref{kinkgas},\ref{K.fixedalpha}) 
has a divergent free energy at zero temperature for $\alpha < 1$
implies the same physics. The precise nature of the ordering 
depends on the details of the band structure and the
intersite interactions. Our results
applies at temperatures above the ordering temperature.

The extension of the self-consistent equations and the kink
gas action to the spinful extended Hubbard model is
straightforward. The form of the scaling equations
implies that our results carry over to the
charge sector of the spin-charge-separated intermediate phase.

After the completion of this work, we learned of the independent 
work of Kajueter and Kotliar\cite{Kajueter} who constructed related
mean field equations in the context of a spinless one-band
fermion model with semi-circular density of states and
found no numerical evidence for non-Fermi liquids in that model.
We thank V. Dobrosavljevic, A. Georges, and A. Sengupta
for stimulating discussions. Q.S. was supported by an A. P. Sloan
Fellowship, and by NSF Grant No. PHY94-07194 at ITP, UCSB.

\figure{The RG flow of the kink-gas action Eqs. 
(\ref{kinkgas}, \ref{K.fixedalpha}) for $\alpha < \sim 2$
and vanishing renormalized symmetry breaking field.
\label{flow}}

\figure{The phase diagram of the Hamiltonian Eq. (\ref{hamiltonian})
in the mixed valence regime. The dashed line is schematic.\label{phases}}

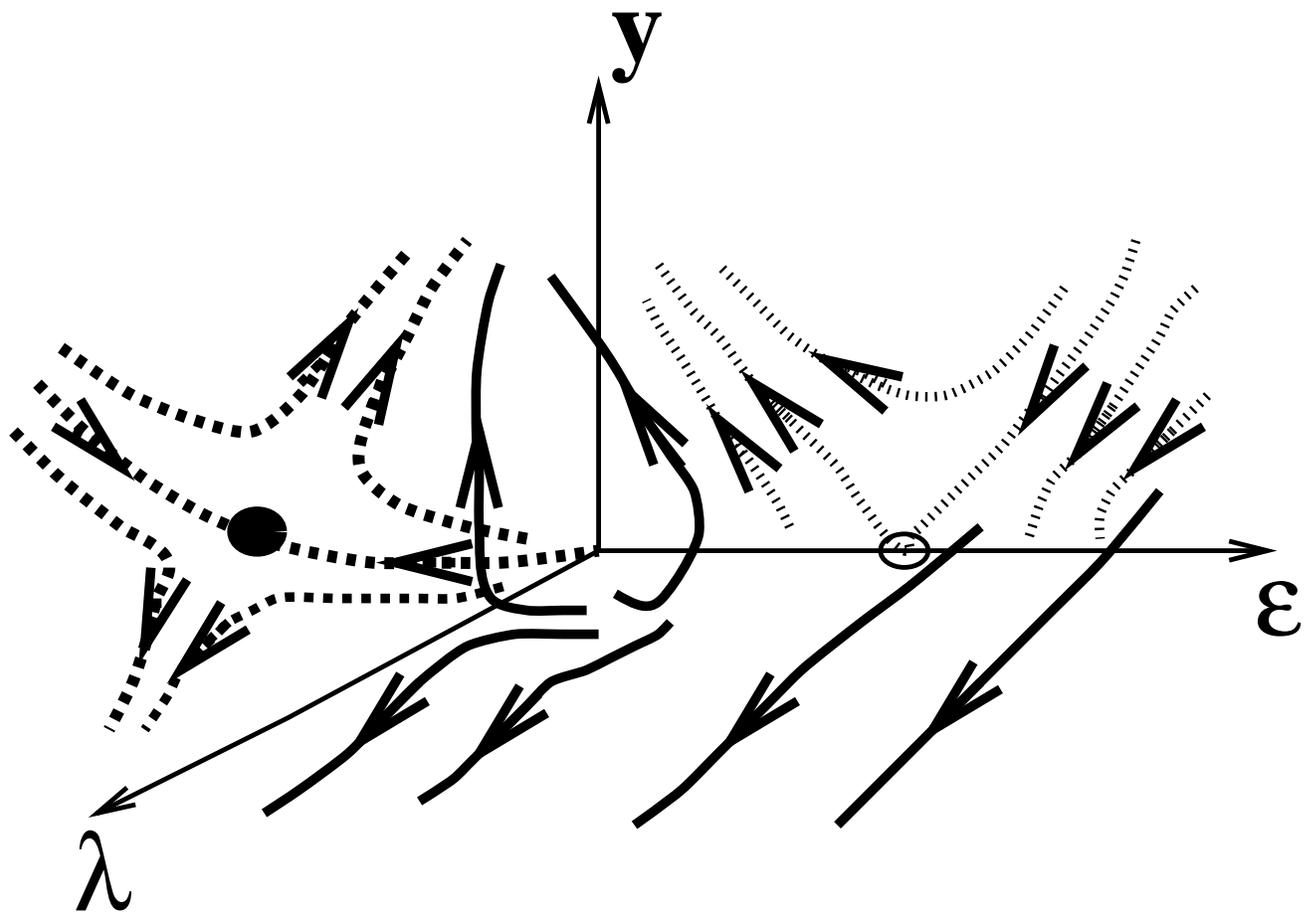
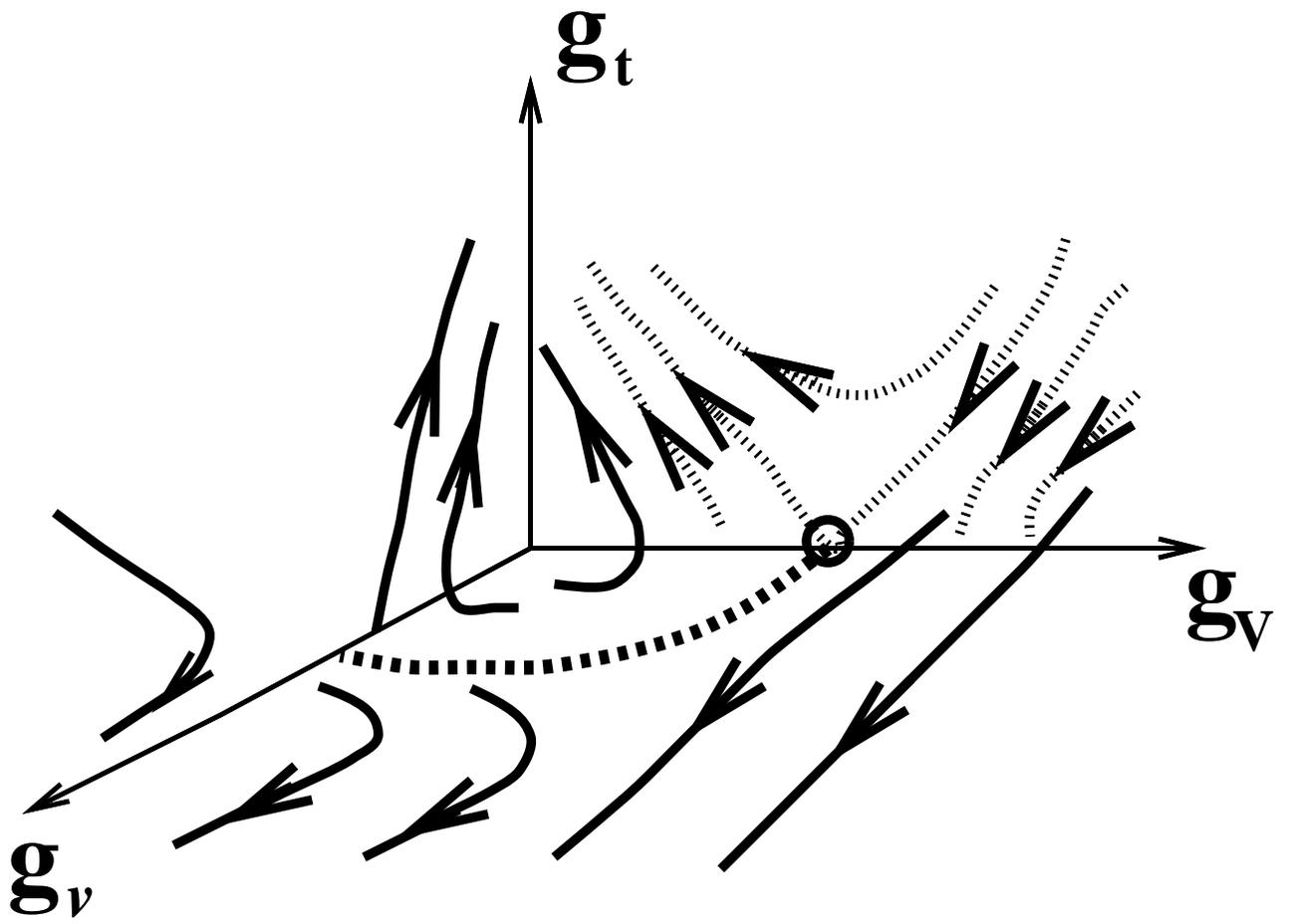
\end{document}